\documentstyle[12pt,worldsci]{article}

\def\thefootnote{\fnsymbol{footnote}}
\def\bea {\begin{eqnarray}}
\def\eea {\end{eqnarray}}
\def\be {\begin{equation}}
\def\ee {\end{equation}}
\def\ben{\begin{enumerate}}
\def\een{\end{enumerate}}
\def\bi{\begin{itemize}}
\def\ei{\end{itemize}}

\def\etal{{\it et al.}}

\def\F{{\cal F}}
\def\A{{\cal A}}

\def\prl {Phys. Rev. Lett.\ }
\def\pl {Phys. Lett.\ }
\def\pr {Phys. Rev.\ }
\def\np {Nucl. Phys.\ }
\def\gA{g_{\mbox{\tiny A}}}

\def\GA{G_{\mbox{\tiny A}}}
\def\GV{G_{\mbox{\tiny V}}}
\def\GF{G_{\mbox{\tiny F}}}
\def\DRV{\Delta_{\mbox{\tiny R}}^{\mbox{\tiny V}}}

\def\mW{m_{\mbox{\tiny W}}}
\def\mP{m_{\mbox{\tiny P}}}
\def\mA{m_{\mbox{\tiny A}}}

\def\mZ{m_{\mbox{\tiny Z}}}

\def\MV{M_{\mbox{\tiny V}}}
\def\thW{\theta_{\mbox{\tiny W}}}

\def\mids{\! \mid \! }

\def\nl{$\,\!$}

\begin{document}

\title{SUPERALLOWED FERMI BETA DECAY -- A STATUS REPORT}
\author{I. S. Towner, E. Hagberg, J. C. Hardy, V.T. Koslowsky and G. Savard \\
{\em AECL, Chalk River Laboratories, Chalk River \\
Ontario K0J 1J0, Canada}}
\maketitle

\begin{abstract}
{\footnotesize
Data on superallowed Fermi beta decay in nuclei are compiled and
electromagnetic corrections discussed. Recommended values for
the weak vector coupling constant and the leading element of the
Cabibbo-Kobayashi-Maskawa mixing matrix are given.  Comments on
neutron decay and limits extracted on extensions to the Standard
Model are made.
}

\end{abstract}

\renewcommand{\thefootnote}{\#\arabic{footnote}}
\setcounter{footnote}{0}

\section{Introduction} \label{intro}

The intensity of superallowed Fermi $\beta$-transitions between
$0^{+},T=1$ nuclear states, as expressed through their $ft$ values, is
expected to be the same for all nuclei.  This statement, based on
the conserved vector current (CVC) hypothesis, follows from two considerations.
First, in the allowed approximation of $\beta$-decay theory,
the Fermi operator being the isospin ladder operator, $\tau_{+}$, leads to
identically-valued matrix elements for all Fermi decays so long
as isospin is an exact symmetry.  Second, CVC ensures the weak vector
coupling constant is not renormalized in a nuclear many-body medium;
it remains a true constant.
However, before these statements
can be tested against experimental data, certain theoretical corrections
have to be applied.  For example, bremsstrahlung processes lead to
radiative corrections, and the breakdown of analogue symmetry by
the presence of charge-dependent forces between nucleons leads to
Coulomb corrections in the nuclear matrix elements.

The nine best-known transitions have reached sufficient experimental
precision that the CVC hypothesis can be tested at the level of a
few parts in $10^4$ and the three-generation Standard Model at the
level of its quantum corrections.  World data on $Q$-values, lifetimes
and branching ratios were thoroughly surveyed \cite{Ha90} in 1989
and updated again \cite{TH95} this year.  The principal new
measurements are:  some $Q$-value determinations by Barker \etal \cite{Ba94a},
an $(n,\gamma)$ and $(p,\gamma)$ $Q$-value difference measurement
\cite{Ki91} \nl , and some lifetime and branching-ratio measurements
\cite{Ko94,Ha94,Sa95} at Chalk River.

Once the CVC test is successfully passed, a value of the weak vector
coupling constant for semi-leptonic decays is determined.  A comparison
with the same quantity for the purely leptonic $\mu$-decay yields
the up-down quark-mixing matrix element, $V_{ud}$, of the
Cabibbo-Kobayashi-Maskawa (CKM) matrix.  The CKM matrix is unitary,
so the sum of the elements in the first row of the matrix should
satisfy

\be
\mids V_{ud} \mids^2 +
\mids V_{us} \mids^2 +
\mids V_{ub} \mids^2 = 1 ,
\label{unit}
\ee

\noindent if the Standard Model for three generations is correct.
The accuracy
with which this relation can be tested is dominated by the accuracy
with which $V_{ud}$ can be determined, which in turn depends on the
accuracy with which the radiative and Coulomb corrections can be
evaluated.

\section{Radiative Corrections} \label{radc}

In the present application we are only interested in the difference
in radiative corrections between nuclear $\beta$-decay and $\mu$-decay,
as only these differences impact on the deduced value of $V_{ud}$.
To first order in the fine-structure constant, the uncorrected
$\beta$-decay rate $\Gamma_{\beta}^0$ is modified \cite{Si78} to

\be
\Gamma_{\beta} = \Gamma_{\beta}^0 \left \{
1 + \frac{\alpha}{2 \pi} \left [ 3 \ln (\mW/\mP) + \overline{g}(E_m)
+ \ln (\mW/\mA) + 2C -4 \ln (\mW/\mZ) + {\A}_g \right ] \right \} ,
\label{gb}
\ee

\noindent where $E_m$ is the maximum electron energy in the $\beta$-decay,
and $\mW$,$\mP$,$\mZ$ are the masses of the $W$-boson, proton and $Z$-boson.
We discuss $\mA$ shortly.  The first two terms in the square brackets
in Eq.\,(\ref{gb}), $3 \ln (\mW/\mP) + \overline{g}(E_m)$, are the
universal photonic contributions arising from the weak vector current
in the $V-A$ theory.  The function $g(E,E_m)$ is defined in Eq.\,(20b)
of Sirlin \cite{Si67} and is here averaged over the electron spectrum.
The next terms, $\ln (\mW/\mA) + 2C$, represent the asymptotic and
nonasymptotic photonic corrections induced by the weak axial-vector
current, which we discuss shortly.  The fifth term arises from
$Z$-exchange graphs, while the sixth term is a small perturbative
QCD correction estimated by Marciano and Sirlin \cite{MS86} to be
${\A}_g = -0.37$.  It is convenient to gather the leading logarithms
together and recast Eq.\,(\ref{gb}) as

\be
\Gamma_{\beta} = \Gamma_{\beta}^0 \left \{
1 + \frac{\alpha}{2 \pi} \left [ 4 \ln (\mZ/\mP) + \overline{g}(E_m)
+ \ln (\mP/\mA) + 2C + {\A}_g \right ] \right \} .
\label{gb1}
\ee

\noindent The first term in the square brackets
gives a universal correction to nuclear
$\beta$-decay of 2.1\%, while the second term contributes a further
1\% but varies slightly from nucleus to nucleus.  The third and fourth
terms have been estimated \cite{MS86} in 1986 to be

\be
\frac{\alpha}{2 \pi} \left [ \ln (\mP/\mA) + 2C \right ]
= (0.12 \pm 0.18) \% ,
\label{a2C}
\ee

\noindent and it is the error here that contributes the principal
error in the evaluation of the radiative correction.  The last term
in Eq.\,(\ref{gb1}) contributes $-0.04\%$ to the $\beta$-decay rate
and is negligible.

Since the first correction term in Eq.\,(\ref{gb1}) is the largest
$O(\alpha)$ correction, Marciano and Sirlin \cite{MS86} have
approximated the effect of higher orders by summing all
leading-logarithmic corrections of $O(\alpha^n \ln^{n} \mZ)$,
$n = 1,2 \cdots$ via a renormalization-group analysis.  Such a
resummation replaces Eq.\,(\ref{gb1}) with

\be
\Gamma_{\beta} = \Gamma_{\beta}^0 \left \{
1 + \frac{\alpha}{2 \pi} \left [ \ln(\mP/\mA) + 2C \right ]
+ \frac{\alpha(\mP)}{2 \pi} \left [ \overline{g}(E_m) + {\A}_g
\right ] \right \} S(\mP,\mZ) ,
\label{gb2}
\ee

\noindent where $S(\mP,\mZ)$ satisfies an appropriate renormalization-group
equation and is evaluated to be 1.0225.  Further, $\alpha(\mu)$ is a
running QED coupling constant having a value at the proton mass of
$\alpha^{-1}(\mP) = 133.93$.  Finally, there are higher-order radiative
corrections of order $Z\alpha^{2}$, $Z^2 \alpha^{3} \cdots$ where
$Z$ is the atomic number of the daughter nucleus in nuclear $\beta$-decay,
which we will denote by $\delta_2$ and $\delta_3$, respectively.
These terms yield \cite{Si87} a correction of order 0.5\% to the
$\beta$-decay rate.

A recent advance has been a closer look at the photonic corrections induced
by the weak axial-vector current, the third and fourth terms in
Eq.\,(\ref{gb}), and the numerical estimates given in Eq.\,(\ref{a2C}).
It is convenient to model these corrections by two graphs.  In the
first, the nucleon emits a $\rho$-meson, which in turn converts to
an $A_1$-meson and a photon.  The $A_1$-meson leads to a weak
axial-vector transition producing an electron-neutrino pair, and
the photon interacts with the electron.  The graph has a closed-loop
integration, comprising the electron, photon, $A_1$-meson and $W$-boson lines,
which is logarithmically divergent in the limit $\mA \rightarrow 0$.
The role of the $A_1$-meson is to provide a low-energy cut-off.  Hence the
factor $\ln (\mW/\mA)$ appears in Eq.\,(\ref{gb}).  Originally Marciano
and Sirlin suggested a conservative range $400 {\rm ~MeV} < \mA <
1600 {\rm ~MeV}$, which spans mass scales from $3 m_{\pi}$ through
the $A_1$ resonance region.  More recently, Sirlin \cite{Si94} has
used a range $\mA /2 < \mA < 2 \mA$ with the central value given
by the $A_1$ resonance mass, $\mA = 1260 {\rm ~MeV}$, a reasonable
procedure since the $A_1$ resonance is now well established \cite{PDG94}
\nl .  We will adopt this latter suggestion for the leading
logarithm in Eq.\,(\ref{a2C}).

The second graph represents the sequential emission from a nucleon
first of a $W$-boson to produce an electron-neutrino pair and
second a photon to be reabsorbed by the electron.  This, together
with the graph with the photon and $W$-boson emitted in reversed order,
are the Born graphs and their contribution is written as $(\alpha /
\pi )C$ in Eq.\,(\ref{gb}).  If the nucleus is taken as a collection
of non-interacting nucleons and the graphs evaluated from a single
nucleon, then a universal contribution \cite{MS86} of $C = 3 \gA
(\mu_p + \mu_n ) I$ is obtained, where I is a loop integral, providing
the axial-vector coupling constant and the nucleon isoscalar magnetic
moment are taken at their free-nucleon values, $\gA = 1.26$ and
$(\mu_p + \mu_n) = 0.88$, respectively.  The integral, $I$, depends
on the choice of form factors introduced at the hadron vertices
required to render the loop integration finite.  Towner \cite{To92}
has studied this dependence and recommends a value $C = 0.881 \pm 0.030$.

An important advance was the observation of Jaus and Rasche \cite{JR90}
that the emission of the $W$-boson and photon do not necessarily have
to be from the same nucleon.  Thus we write

\be
C = C_{\rm Born} + C_{NS} ,
\label{CNS}
\ee

\noindent where the departure of the value of C from the
universal single-nucleon
value of $C_{\rm Born} = 0.881 \pm 0.030$ is attributed to both the
nuclear-structure dependent two-nucleon graphs and a quenching in
the single-nucleon graphs due to the renormalization of the coupling
constants $\gA$ and $(\mu_p + \mu_n)$ in the nuclear medium.  Two-nucleon
graph calculations are to be found in refs. \cite{To92,BBJR92} and
quenching results in ref.\cite{To94} \nl .  With the exception of
$^{10}$C and $^{14}$O, the correction is generally small,
$(\alpha / \pi )C_{NS}$ being less than $0.09\%$.  For the ensuing
analysis, we adopt the values from Towner \cite{To94} \nl .

It has been convenient in the past to separate the radiative correction
into those terms that are nuclear-structure dependent, $\delta_R$,
and those that are not, $\DRV$.  We will continue with this separation,
although it is cumbersome with the term $\overline{g}(E_m)$ contained
in the renormalization-group extrapolation.  Thus we define

\be
\Gamma_{\beta} = \Gamma_{\beta}^0 (1 + \delta_R )(1 + \DRV) ,
\label{gb3}
\ee

\noindent with

\bea
1 + \delta_R & = & 1 + \frac{\alpha}{2 \pi} \left [
\overline{g}(E_m) + \delta_2 + \delta_3 + 2 C_{NS} \right ] ,
\label{dR} \\
1 + \DRV & = & \left \{ 1 + \frac{\alpha}{2 \pi} \left [
\ln (\mP/\mA) + 2 C_{\rm Born} \right ] + \frac{\alpha(\mP)}{2 \pi}
\left [ \overline{g}(E_m) + {\A}_g \right ] \right \}
\nonumber \\
& &~~~~~~ \times S(\mP,\mZ) \left \{ 1 - \frac{\alpha}{2 \pi}
\overline{g}(E_m) \right \} .
\label{DR}
\eea

\noindent  We persevere with this separation in Eq.\,(\ref{gb3}) because
only the nucleus-dependent radiative corrections are required for the
testing of the CVC hypothesis through the constancy of the $ft$ values.
The values of $\delta_R$ adopted for this test are recorded in Table
\ref{tab1}.  For the nucleus-independent correction, we obtain

\be
\DRV = (2.40 \pm 0.08) \% ,
\label{DRvalu}
\ee

\noindent where the error reflects the range adopted for $\mA$ for
the low-energy cut-off.

\section{Coulomb Corrections} \label{coulc}

Both Coulomb and charge-dependent nuclear forces destroy isospin
symmetry and reduce the value of the Fermi matrix element between
analogue states from its simple value of $\langle \MV \rangle^2
= 2$ to a corrected value of $2 (1 - \delta_C)$.
Two physical phenomena can contribute to $\delta_C$. First, the
degree of configuration mixing in the shell-model wavefunctions varies from
member to member within an isospin multiplet, leading to a correction
$\delta_{IM}$.  Second, protons are typically less bound than neutrons;
thus the tail of the radial wavefunction for protons extends further
than that for neutrons, so the radial overlap of the parent and
daughter nucleus is reduced from the normally assumed value of unity --
a correction $\delta_{RO}$. Generally, $\delta_{RO}$ is larger than
$\delta_{IM}$.

There are two comprehensive sets of calculations of these corrections:
by Towner, Hardy and Harvey \cite{THH77,To89,Ha94} (THH), and by
Ormand and Brown \cite{OB89,OB95} \nl . Their relative merits are
discussed in ref. \cite{Ha90} and will not be repeated here.
Ormand and Brown have just released a revised calculation of $\delta_C$,
and we will adopt these values in what follows.  A close examination
indicates that the two sets exhibit very similar nucleus-to-nucleus
variations:  both predict the largest correction for $^{38m}$K and the
smallest for $^{10}$C.  However, there is a systematic difference,
with the THH values being larger than the OB values by 0.08\%
on the average.
Accordingly, we adopt for $\delta_C$ of each transition the unweighted
average of the two independent calculations for that transition and
then analyze the scatter of all nine pairs of $\delta_C$ calculations
about their respective averages to obtain a standard deviation of
$\pm 0.03 \%$ for the purposes of
the CVC test.  After the CVC test has been satisfied and data on the nine
nuclear transitions are averaged together, a further ``systematic" error of
$\pm 0.04 \%$ is incorporated to represent the systematic difference
between the two calculations.  The $\delta_C$ corrections are listed
in Table \ref{tab1}.

\begin{table}[t]
\begin{center}
\caption{$ft$-values, theoretical corrections $\delta_R$ and
$\delta_C$, and the corrected ${\cal F} t$-values for the nine
accurately-measured superallowed Fermi $\beta$-transitions.
\label{tab1}}
\vskip 1mm
\begin{tabular}{lllll}
\hline \\[-3mm]
 & \multicolumn{1}{c}{$ft$} & $\delta_R(\%)$ & $\delta_C(\%)^{a}$ &
\multicolumn{1}{l}{~~~$\F t$} \\
\hline \\[-3mm]
$^{10}$C & 3040.1(51) & 1.30(4) & 0.16(3) & 3074.4(54) \\
$^{14}$O & 3038.1(18) & 1.26(5) & 0.22(3) & 3069.7(26) \\
$^{26m}$Al & 3035.8(17) & 1.45(2) & 0.31(3) & 3070.0(21) \\
$^{34}$Cl & 3048.4(19) & 1.33(3) & 0.61(3) & 3070.1(24) \\
$^{38m}$K & 3047.9(26) & 1.33(4) & 0.62(3) & 3069.4(31) \\
$^{42}$Sc & 3045.1(14) & 1.47(5) & 0.41(3) & 3077.3(23) \\
$^{46}$V & 3044.6(18) & 1.40(6) & 0.41(3) & 3074.4(27) \\
$^{50}$Mn & 3043.7(16) & 1.40(7) & 0.41(3) & 3073.8(27) \\
$^{54}$Co & 3045.8(11) & 1.39(7) & 0.52(3) & 3072.2(27) \\
\hline \\
\multicolumn{5}{l}{{\small $^{a}$Tabulated uncertainties represent
``statistical" scatter only.}}
\end{tabular}
\end{center}
\end{table}

Three recent developments can have a bearing on the $\delta_C$
calculations: \newline
{\it (a)}  Wilkinson \cite{Wi90} notes that the experimental transition
rates, when corrected for radiative and Coulomb effects, still display
a weak dependence on the atomic number, $Z$, of the daughter nucleus,
which he attributes to some inadequacy in the Coulomb-correction calculation.
His largely empirical approach to this is to remove the nucleus-to-nucleus
fluctations evident in both the THH and OB calculations from the experimental
$ft$ values and then fit a smooth curve through the resultant values, which
can be extrapolated back to $Z = 0$ to obtain a recommended value. It
should be remembered, however, that this is a heuristic procedure,
not based on an {\it ab initio} calculation.  Initially \cite{Wi90} \nl ,
it was thought that the inadequacy in the $\delta_C$ calculation might
be an inadequate allowance for the effects of the spectator nucleons
(the core nucleons in shell-model approaches) in the radial overlaps,
but recent analyses \cite{Wi95} show such effects are negligible.

\noindent {\it (b)} Barker \cite{Ba94b} has computed a full set of
$\delta_C$ values in an $R$-matrix formulation in which the configuration
space is divided into internal and external regions. This produces
an obvious dependence on the choice of channel radius, $a$;
smaller $\delta_C$ values arise with larger values of $a$.
Since physically-motivated values of $a$ increase with nuclear
mass number as $A^{1/3}$, small $\delta_C$ values are obtained in
heavier nuclei.  In addition, the computations produce
overall small values with $\delta_C < 0.2 \%$ in all cases
and $< 0.05 \%$ in the $fp$-shell nuclei $^{46}$V, $^{50}$Mn and $^{54}$Co.
The results lead to corrected $ft$ values that are not consistent with
the CVC hypothesis.

\noindent {\it (c)}  In a recent preprint, Sagawa, van Giai and
Suzuki \cite{SGS95} compute the isospin mixing in the ground states of
several odd-odd
$N = Z$ nuclei in the Hartree-Fock approximation taking into
account forces that break charge symmetry and charge independence.
It must be stressed that isospin mixing in ground states is not the
same as analogue-symmetry breaking in Fermi beta decay:
the latter depends on the difference in isospin mixing in
adjacent nuclei.  Sagawa \etal \  find considerable isospin mixing
arising from the core nucleons, and obtain values as large as $1\%$
for $^{54}$Co,  which they cite as a warning that considerable
core effects may be important in $\delta_C$ calculations.

\section{The CVC test and the effective vector coupling constant} \label{cvcs}

\begin{figure}[t]
\begin{center}
\setlength{\unitlength}{0.240900pt}
\ifx\plotpoint\undefined\newsavebox{\plotpoint}\fi
\sbox{\plotpoint}{\rule[-0.175pt]{0.350pt}{0.350pt}}%
\begin{picture}(1500,900)(0,0)
\tenrm
\sbox{\plotpoint}{\rule[-0.175pt]{0.350pt}{0.350pt}}%
\put(264,315){\rule[-0.175pt]{4.818pt}{0.350pt}}
\put(242,315){\makebox(0,0)[r]{3070}}
\put(1416,315){\rule[-0.175pt]{4.818pt}{0.350pt}}
\put(264,473){\rule[-0.175pt]{4.818pt}{0.350pt}}
\put(242,473){\makebox(0,0)[r]{3075}}
\put(1416,473){\rule[-0.175pt]{4.818pt}{0.350pt}}
\put(264,630){\rule[-0.175pt]{4.818pt}{0.350pt}}
\put(242,630){\makebox(0,0)[r]{3080}}
\put(1416,630){\rule[-0.175pt]{4.818pt}{0.350pt}}
\put(354,158){\rule[-0.175pt]{0.350pt}{4.818pt}}
\put(354,113){\makebox(0,0){5}}
\put(354,767){\rule[-0.175pt]{0.350pt}{4.818pt}}
\put(580,158){\rule[-0.175pt]{0.350pt}{4.818pt}}
\put(580,113){\makebox(0,0){10}}
\put(580,767){\rule[-0.175pt]{0.350pt}{4.818pt}}
\put(805,158){\rule[-0.175pt]{0.350pt}{4.818pt}}
\put(805,113){\makebox(0,0){15}}
\put(805,767){\rule[-0.175pt]{0.350pt}{4.818pt}}
\put(1030,158){\rule[-0.175pt]{0.350pt}{4.818pt}}
\put(1030,113){\makebox(0,0){20}}
\put(1030,767){\rule[-0.175pt]{0.350pt}{4.818pt}}
\put(1256,158){\rule[-0.175pt]{0.350pt}{4.818pt}}
\put(1256,113){\makebox(0,0){25}}
\put(1256,767){\rule[-0.175pt]{0.350pt}{4.818pt}}
\put(264,158){\rule[-0.175pt]{282.335pt}{0.350pt}}
\put(1436,158){\rule[-0.175pt]{0.350pt}{151.526pt}}
\put(264,787){\rule[-0.175pt]{282.335pt}{0.350pt}}
\put(45,562){\makebox(0,0)[l]{\shortstack{$\F t(s)$}}}
\put(850,23){\makebox(0,0){$Z$ of daughter}}
\put(264,158){\rule[-0.175pt]{0.350pt}{151.526pt}}
\put(354,454){\circle*{24}}
\put(444,306){\circle*{24}}
\put(670,315){\circle*{24}}
\put(850,318){\circle*{24}}
\put(940,296){\circle*{24}}
\put(1030,545){\circle*{24}}
\put(1120,454){\circle*{24}}
\put(1211,435){\circle*{24}}
\put(1301,384){\circle*{24}}
\put(354,284){\rule[-0.175pt]{0.350pt}{81.665pt}}
\put(344,284){\rule[-0.175pt]{4.818pt}{0.350pt}}
\put(344,623){\rule[-0.175pt]{4.818pt}{0.350pt}}
\put(444,224){\rule[-0.175pt]{0.350pt}{39.508pt}}
\put(434,224){\rule[-0.175pt]{4.818pt}{0.350pt}}
\put(434,388){\rule[-0.175pt]{4.818pt}{0.350pt}}
\put(670,249){\rule[-0.175pt]{0.350pt}{31.799pt}}
\put(660,249){\rule[-0.175pt]{4.818pt}{0.350pt}}
\put(660,381){\rule[-0.175pt]{4.818pt}{0.350pt}}
\put(850,243){\rule[-0.175pt]{0.350pt}{36.376pt}}
\put(840,243){\rule[-0.175pt]{4.818pt}{0.350pt}}
\put(840,394){\rule[-0.175pt]{4.818pt}{0.350pt}}
\put(940,199){\rule[-0.175pt]{0.350pt}{46.975pt}}
\put(930,199){\rule[-0.175pt]{4.818pt}{0.350pt}}
\put(930,394){\rule[-0.175pt]{4.818pt}{0.350pt}}
\put(1030,473){\rule[-0.175pt]{0.350pt}{34.690pt}}
\put(1020,473){\rule[-0.175pt]{4.818pt}{0.350pt}}
\put(1020,617){\rule[-0.175pt]{4.818pt}{0.350pt}}
\put(1120,369){\rule[-0.175pt]{0.350pt}{40.953pt}}
\put(1110,369){\rule[-0.175pt]{4.818pt}{0.350pt}}
\put(1110,539){\rule[-0.175pt]{4.818pt}{0.350pt}}
\put(1211,350){\rule[-0.175pt]{0.350pt}{40.953pt}}
\put(1201,350){\rule[-0.175pt]{4.818pt}{0.350pt}}
\put(1201,520){\rule[-0.175pt]{4.818pt}{0.350pt}}
\put(1301,300){\rule[-0.175pt]{0.350pt}{40.712pt}}
\put(1291,300){\rule[-0.175pt]{4.818pt}{0.350pt}}
\put(1291,469){\rule[-0.175pt]{4.818pt}{0.350pt}}
\sbox{\plotpoint}{\rule[-0.350pt]{0.700pt}{0.700pt}}%
\put(264,388){\usebox{\plotpoint}}
\put(264,388){\rule[-0.350pt]{282.335pt}{0.700pt}}
\end{picture}
\vskip 1mm
\caption{$\F t$ values for the nine precision data and the best
least squares one-parameter fit  \label{fig1}}
\end{center}
\end{figure}
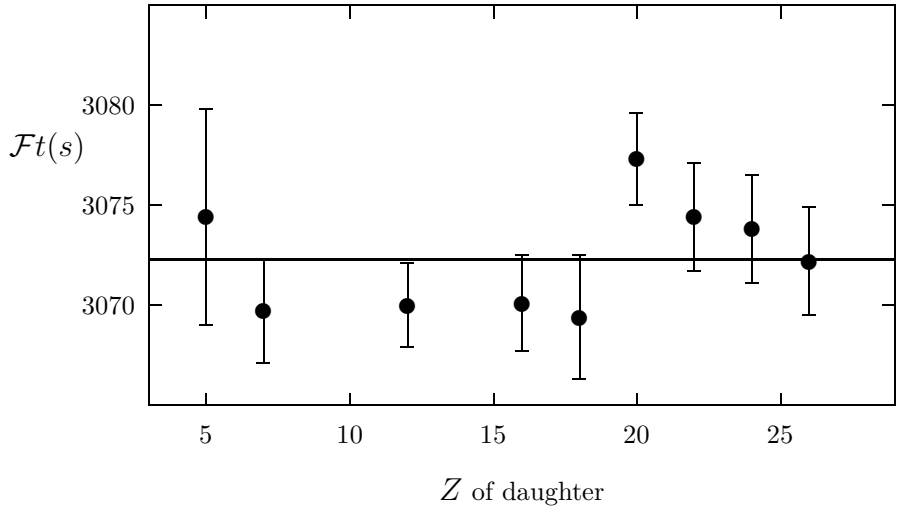

The test of the conserved vector current (CVC) hypothesis is that the
$ft$ values for superallowed beta decay, when suitably corrected for
electromagnetic effects as just discussed, should all be constant.
We write

\bea
ft (1 + \delta_R) (1 - \delta_C) & \equiv & \F t = K/(2 \GV^{\prime 2}),
\label{Ft} \\
K & = & 2 \pi^3 \ln 2 \hbar^7 / (m_e^5 c^4) ,
\label{Kvalu} \\
\GV^{\prime 2} & = & \GV^2 ( 1 + \DRV ) ,
\label{GVp}
\eea

\noindent where $\GV^{\prime}$ is the effective vector coupling
constant.  It includes the nucleus-independent radiative correction,
$\DRV$, which is not relevant for the test of CVC.  The status of the
experimental data and the theoretical corrections $\delta_R$ and
$\delta_C$ are summarised in Table \ref{tab1}.  The weighted
average of the $\F t$ values (with statistical uncertainty only) is
$3072.3 \pm 1.0$ s with a corresponding $\chi^2$ per degree of
freedom of 1.20.  A two-parameter fit to the function, $\F t =
\F t(0) \left [ 1 + a_1 Z \right ]$, gives only a marginal
decrease in the $\chi^2/\nu$ to 1.00 with the parameter $a_1 =
(0.7 \pm 0.5) \times 10^{-4}$.  Statistically there is
insufficient evidence for a residual $Z$-dependence.  The CVC
hypothesis is therefore verified at the level of $4 \times 10^{-4}$.
The nine $\F t$ values are displayed in Fig. \ref{fig1}.

Having substantiated the consistency of the experimental $\F t$
values, we now turn to the extraction of the effective nuclear vector
coupling constant, $\GV^{\prime}$, which is directly related by
Eq.\,(\ref{Ft}) to the average $\F t$ value.  It is important now,
however, to include in the $\F t$-value uncertainty the effect of
``systematic" as well as ``statistical" uncertainties in the
charge-correction calculation, as described in Sect.\ref{coulc}.
These ``systematic" or model-dependent effects were estimated to be $\pm
0.04\%$. Thus the average $\F t$ value (with both ``statistical"
and ``systematic" errors included) becomes

\be
\F t = 3072.3 \pm 2.0 ~{\rm s} ,
\label{Ftavg}
\ee

\noindent and so

\be
\GV^{\prime}/(\hbar c)^3 = (1.14959 \pm 0.00038) \times 10^{-5}
{}~{\rm GeV}^{-2} .
\label{GVpr}
\ee

\noindent The error bars in Eqs.\,(\ref{Ftavg}) and (\ref{GVpr})
are dominated by
the theoretical uncertainties in $\delta_C$.

\section{The value of $V_{ud}$ and the unitarity test} \label{Vudut}

The superallowed nuclear beta transitions are examples of semi-leptonic
weak decays, which, when compared with the pure leptonic muon decay,
yield an experimental value for $V_{ud}$ via the relationship

\be
\GV^{\prime 2} / \GF^2 = V_{ud}^2 (1 + \DRV) .
\label{Vud2}
\ee

\noindent Here, $\DRV$ is the nucleus-independent radiative correction
given in Eq.\,(\ref{DRvalu}). Thus with

\be
\GF /(\hbar c)^3 = (1.166387 \pm 0.000021) \times 10^{-5}
{}~{\rm GeV}^{-2} ,
\label{GFvalu}
\ee

\noindent derived from muon decay \cite{PDG94} \nl , we obtain

\be
V_{ud} = 0.9740 \pm 0.0005 .
\label{Vudvalu}
\ee

\noindent The quoted uncertainty is dominated by uncertainties in the
theoretical corrections, $\DRV$ and $\delta_C$; experimental errors on
the input data for the $\F t$-value contribute less than $\pm 0.0002$
to the uncertainty in Eq.\,(\ref{Vudvalu}).

Evidently, $V_{ud}$ is by far the largest of the three matrix elements
needed for the unitarity test, Eq.\,(\ref{unit}), and its uncertainty
is the most crucial.  The value of $V_{us}$ is taken to be \cite{PDG94}
$V_{us} = 0.2205 \pm 0.0018$, the average of two results, one from
the analysis of $K_{e3}$ decays, the other from hyperon decays.
Finally, $V_{ub}$ can be derived, with substantial uncertainty, from
the semi-leptonic decay of $B$ mesons, but is sufficiently
small \cite{PDG94} \nl , at $0.0032 \pm 0.0009$, as to be negligible
in the unitarity test.  The result,

\be
\mids V_{ud} \mids^2 +
\mids V_{us} \mids^2 +
\mids V_{ub} \mids^2 = 0.9972 \pm 0.0013 ,
\label{utnuc}
\ee

\noindent indicates a violation of the unitarity condition for three
generations by more than twice the estimated error.

\section{Neutron decay} \label{neutd}

Because of the complexities involved in finite nuclei, we briefly
turn our attention to the study of neutron decay, where
nuclear-structure corrections can be expected to be minimal and
$\delta_C$ is essentially zero.  There have been impressive advances in
the measurement of the neutron mean lifetime in the past six years;
the average \cite{TH95} of eleven measurements is
$\tau_{n} = 887.0 \pm 2.0$ s. The problem for the extraction of the weak
vector coupling constant and hence $V_{ud}$ in neutron decay
is that the transition is a mix of vector and axial-vector
components and a separate $\beta$-asymmetry experiment is required to
separate the two.  There are three recent accurate $\beta$-asymmetry
experiments, not completely in accord; their weighted average
\cite{TH95} leads to $\lambda \equiv \GA^{\prime}/\GV^{\prime} =
-1.2599 \pm 0.0025$.  From these two data, the individual coupling
constants can be deduced:

\bea
\GA^{\prime} /(\hbar c)^3 & = & - (1.4566 \pm 0.0018) \times 10^{-5}
{}~~ {\rm GeV}^{-2} ,
\label{gGA}  \\
\GV^{\prime} /(\hbar c)^3 & = &  (1.1561 \pm 0.0023) \times 10^{-5}
{}~~ {\rm GeV}^{-2}  ~~~~ [{\rm neutron}] .
\label{GVneut}
\eea

\noindent Note the neutron result for $\GV^{\prime}$ is considerably
less precise
but, even so, it is in serious disagreement with nuclear decays,
Eq.\,(\ref{GVpr}).

With the use of the nucleus-independent radiative correction,
$\DRV$ of Eq.\,(\ref{DRvalu}), the neutron result
for the vector coupling constant can be used to test the unitarity
of the CKM matrix, yielding

\be
\mids V_{ud} \mids ^{2} +
\mids V_{us} \mids ^{2} +
\mids V_{ub} \mids ^{2} = 1.0082 \pm 0.0040,
\label{unitn}
\ee

\noindent which differs from unitarity,
but in the opposite sense to the result from
nuclear $\beta$ decay.  In the absence of any viable explanation
within the Standard Model for the discrepancy between the neutron and
nuclear results, nor any reason to prefer one over the other, we
can do no better than take the weighted average of both results
for $\GV^{\prime}$, suitably inflating the resultant uncertainty.
This leads to

\bea
\GV^{\prime}/(\hbar c)^3   & = &  (1.1498 \pm 0.0010) \times 10^{-5}
{}~~ {\rm GeV}^{-2}  ~~~~ [{\rm average}] ,
\nonumber  \\
\mids V_{ud} \mids & = & 0.9741 \pm 0.0010  ~~~~ [{\rm average}] ,
\label{GVavg}
\eea

\noindent and

\be
\mids V_{ud} \mids ^{2} +
\mids V_{us} \mids ^{2} +
\mids V_{ub} \mids ^{2}  =  0.9975 \pm 0.0020 ~~~~ [{\rm average}],
\label{unitavg}
\ee

\noindent which does not disagree significantly from unitarity.

\section{Beyond the Standard Model}
\label{bsm}

The failure of the experimental test to reproduce the unitarity result
can be used to set limits on extensions to the Standard Model.  Two
examples follow:

\noindent {\it (a)} The existence of additional neutral gauge bosons,
beyond the usual $\gamma$ and $Z$ of the Standard
$SU(2)_L \times U(1)$ Model would signal the presence of additional
$U(1)$ groups and would lead to further contributions to the
radiative correction $\DRV$.  This additional contribution could then
be responsible for the failure of the experimental test of unitarity.
Marciano and Sirlin \cite{MS87} give a quantitative estimate of this
for the grand unified models of the type $SO(10) \rightarrow
SU(3)_c \times SU(2)_L \times U(1) \times U(1)_{\chi}$.  Writing

\be
\mids V_{ud} \mids ^{2} +
\mids V_{us} \mids ^{2} +
\mids V_{ub} \mids ^{2}  =  1 - \Delta ,
\label{unitD}
\ee

\noindent then for the grand unified model, $SO(10)$, Marciano and
Sirlin \cite{MS87} obtain

\be
\Delta = - \frac{3 \alpha}{2 \pi \cos^2 \thW } \frac{\ln x}
{x - 1} ,
\label{DSO10}
\ee

\noindent where $x = (m_{Z_{\chi}}/\mW)^2$, $\sin^2 \thW = 0.23$
and $\alpha$ is the fine-structure constant. Thus, on the one hand,
with $\Delta$
positive from Eq.\,(\ref{utnuc}), the correction has the wrong sign
to provide a limit on the $Z_{\chi}$ boson mass, while, on the other
hand, the
neutron data in  Eq.\,(\ref{unitn}) gives limits of
$ 30 < m_{Z_{\chi}} < 85$ GeV, which are unacceptable, since experiments
\cite{PDG94,MS87} searching for departures from the Standard Model
in neutrino-electron scattering or searching directly in
$p \overline{p}$ colliders have concluded that $ m_{Z_{\chi}} >
320$ GeV.

\noindent {\it (b)} Another example is an extension of the Standard
Model into a left-right symmetric form, $SU(2)_L \times SU(2)_R
\times U(1)$, where the $W$-boson associated with the group
$SU(2)_R$, ($W_R$), is deemed to be much heavier than the traditional
$W$-boson associated with $SU(2)_L$, ($W_L$), such that right-hand
couplings in weak interactions are suppressed.  The mass eigenstates
for the two bosons will not necessarily coincide with the weak-interaction
eigenstates, but will be a linear combination of them,

\bea
W_1 & = & W_L \cos \zeta - W_R \sin \zeta ,
\nonumber \\
W_2 & = & W_L \sin \zeta + W_R \cos \zeta ,
\label{Wmix}
\eea

\noindent where $\zeta$ is known as the left-right mixing angle.  The
presence of right-hand currents will lead to a correction to the value
of $V_{ud}$ determined from $\beta$-decay under the assumption of
left-hand currents only.  If we assume that the failure of the
unitarity test in Eq.\,(\ref{utnuc}) is entirely due to the presence
of right-hand currents, then the value for the mixing angle, $\zeta$,
is obtained from the expression

\be
\Delta = 2 \zeta .
\label{Dzeta}
\ee

\noindent The result from superallowed beta decay
for the mixing angle is $\zeta = 0.0014 \pm
0.0006$, which is small but non-zero.  This is a much tighter limit
than can be obtained \cite{Ca92} from other nuclear physics
experiments.

\section{Conclusions}
\label{concl}

In summary, we note that:

\noindent {\it (a)} The average $\F t$ value of Eq.\,(\ref{Ftavg}) for
nine precision data with a $\chi^2$ per degree of freedom of 1.20
provides strong confirmation of the CVC hypothesis to four parts
in $10^4$.

\noindent {\it (b)} There is a serious discrepancy between $\GV^{\prime}$
determined from superallowed beta decay and from neutron decay,
which has yet to be resolved.

\noindent {\it(c)} The significance of the failure of the unitarity test,
Eq.\,(\ref{utnuc}), is not yet settled.  It may, of course, indicate the
need for some extension to the three-generation Standard Model.
However, it may also reflect some undiagnosed inadequacy in the evaluation
of $V_{us}$ (as already suggested for other reasons in ref.\cite{Ga92})
or possibly in the $\delta_C$ corrections used to determine $V_{ud}$.
There is little scope left in the data in Fig.\ref{fig1} for
introducing significant additional $Z$-dependence in $\delta_C$,
as has been suggested recently \cite{Wi90} \nl .


\begin{thebibliography} {999}

\bibitem{Ha90}
J.C. Hardy, I.S. Towner, V.T. Koslowsky, E. Hagberg and H. Schmeing,
\np {\bf A509}, 429 (1990)

\bibitem{TH95}
I.S. Towner and J.C. Hardy, in {\em The Nucleus as a Laboratory for
Studying Symmetries and Fundamental Interactions}, eds. E.M. Henley
and W.C. Haxton (World-Scientific, Singapore, 1995)
to be published

\bibitem{Ba94a}
S.A. Brindhaban and P.H. Barker, \pr {\bf C49}, 2401 (1994);
S. Lin, S.A. Brindhaban and P.H. Barker, \pr {\bf C49}, 3098 (1994);
P.A. Amundsen and P.H. Barker, \pr {\bf C50}, 2466 (1994)

\bibitem{Ki91}
S.W. Kikstra, Z. Guo, C. van der Leun, P.M. Endt, S. Raman, T.A. Walkiewicz,
J.W. Starner, E.T. Jueney and I.S. Towner,
\np {\bf A529}, 39 (1991);
T.A. Walkiewicz, S. Raman, E.T. Jurney, J.W. Starner and J.E. Lynn,
\pr {\bf C45}, 1597 (1992)

\bibitem{Ko94}
V.T. Koslowsky \etal, to be published;  preliminary results appear in
E. Hagberg, V.T. Koslowsky, I.S. Towner, J.G. Hykawy, G. Savard, T. Shinozuka,
P.P. Unger and H. Schmeing, {\it Nuclei far from Stability/Atomic Masses
and Fundamental Constants} (Institute of Physics Conference Series \# 132,
ed. R. Neugart and A. W\"{o}hr, 1994) p.783

\bibitem{Ha94}
E. Hagberg, V.T. Koslowsky, J.C. Hardy, I.S. Towner, J.G. Hykawy, G. Savard
and T. Shinozuka, \prl {\bf 73}, 396 (1994)

\bibitem{Sa95}
G. Savard, A. Galindo-Uribarri, E. Hagberg, J.C. Hardy, V.T. Koslowsky,
D.C. Radford
and I.S. Towner, \prl {\bf 74}, 1521 (1995)

\bibitem{Si78}
A. Sirlin, Rev. Mod. Phys. {\bf 50}, 573 (1978)

\bibitem{Si67}
A. Sirlin, \pr {\bf 164}, 1767 (1967)

\bibitem{MS86}
W.J. Marciano and A. Sirlin, \prl {\bf 56}, 22 (1986)

\bibitem{Si87}
A. Sirlin, \pr {\bf D35}, 3423 (1987);
W. Jaus and G. Rasche, \pr {\bf D35}, 3420 (1987);
A. Sirlin and R. Zucchini, \prl {\bf 57}, 1994 (1986)

\bibitem{Si94}
A. Sirlin, in {\em Precision Tests of the Standard
Electroweak Model} ed. P. Langacker (World-Scientific, Singapore, 1994)
to be published

\bibitem{PDG94}
Particle Data Group, \pr {\bf D50}, 1173 (1994)

\bibitem{To92}
I.S. Towner, \np {\bf A540}, 478 (1992)

\bibitem{JR90}
W. Jaus and G. Rasche, \pr {\bf D41}, 166 (1990)

\bibitem{BBJR92}
F.C. Barker, B.A. Brown, W. Jaus and G. Rasche, \np {\bf A540}, 501
(1992)

\bibitem{To94}
I.S. Towner, \pl {\bf B333}, 13 (1994)

\bibitem{THH77}
I.S. Towner, J.C. Hardy and M. Harvey, \np {\bf A284}, 269 (1977)

\bibitem{To89}
I.S. Towner, in {\em Symmetry Violations in Subatomic Physics}, eds.
B. Castel and P.J. O'Donnel (World-Scientific, Singapore, 1989) p.211

\bibitem{OB89}
W.E. Ormand and B.A. Brown, \prl {\bf 62}, 866 (1989); \np {\bf A440},
274 (1985)

\bibitem{OB95}
W.E. Ormand and B.A. Brown, {\em Isospin-mixing corrections for $fp$-shell
Fermi transitions}, preprint, to be published

\bibitem{Wi90}
D.H. Wilkinson, \np {\bf A511}, 301 (1990); Nucl. Inst. and Method
{\bf 335}, 172,182,201 (1993); Zeit. Phys. {\bf A348}, 129 (1994)

\bibitem{Wi95}
D.H. Wilkinson, \np {\bf A587}, 421 (1995)

\bibitem{Ba94b}
F.C. Barker, \np {\bf A579}, 62 (1994);
F.C. Barker, \np {\bf A537}, 143 (1992)

\bibitem{SGS95}
H. Sagawa, N. van Giai and T. Suzuki, {\em Isospin mixing and sum
rule of superallowed Fermi $\beta$ decay}, preprint, to be published

\bibitem{MS87}
W.J. Marciano and A. Sirlin, \pr {\bf D35}, 1672 (1987);
P.Langacker and M Luo, \pr {\bf D45}, 278 (1992)

\bibitem{Ca92}
A.S. Carnoy, J. Deutsch, R. Prieels, N. Severijns and P.A. Quin,
J. Phys. {\bf G18}, 823 (1992);
A.S. Carnoy, J. Deutsch, T.A. Girard and R. Prieels, \prl {\bf 65}, 3249
(1990) and \pr {\bf C43}, 2825 (1991)

\bibitem{Ga92}
A. Garcia, R. Huerta and P. Kielanowski, \pr {\bf D45}, 879 (1992)
\end{thebibliography}
\end{document}